\documentclass[12pt,a4paper,UTF8]{article}
\usepackage{amsmath,mathtools,amsfonts,amsmath,amssymb,mathrsfs,dsfont,cases,amsthm,bm,bbm}
\usepackage{color,xcolor}
\usepackage{graphicx,subfigure,threeparttable}
\usepackage{titlesec}
\usepackage{geometry}
\usepackage{natbib}
\usepackage[colorlinks=true,citecolor=blue,urlcolor=blue,linkcolor=blue]{hyperref}%
\usepackage{array,booktabs,makecell,tabularx,multirow,diagbox}  
\usepackage{rotating}

\newtheoremstyle{thry}
{1em}
{1em}
{\itshape\rmfamily}
{}
{\scshape\large}
{.}
{.5em}
{}
\theoremstyle{thry}

\newtheorem{Thm}{\indent Theorem}[section]
\newtheorem{Lem}{\indent Lemma}[section]

\numberwithin{equation}{section}

\title{Variance-reduced sampling importance resampling}
\author{%
{Yao Xiao$^{1}$, Kang Fu$^{2}$, and Kun Li$^{3}$\thanks{Corresponding author: likun\_post@163.com.}}
\vspace{1.6mm}\\
\fontsize{11}{10}\selectfont\itshape
$^1$\, School of Mathematics and Statistics, Beijing Institute of Technology, Beijing, China. \\
\fontsize{11}{10}\selectfont\itshape
$^2$\, School of Mathematics and Statistics, Central China Normal University, Wuhan, China. \\
\fontsize{11}{10}\selectfont\itshape
$^3$\, School of Statistics and Mathematics, Zhongnan University of Economics and Law, Wuhan, China}
\date{\empty}

\begin{document}
\maketitle

\begin{abstract}
	The sampling importance resampling method is widely utilized in various fields, such as numerical integration and statistical simulation. In this paper, two modified methods are presented by incorporating two variance reduction techniques commonly used in Monte Carlo simulation, namely antithetic sampling and Latin hypercube sampling, into the process of sampling importance resampling method respectively. Theoretical evidence is provided to demonstrate that the proposed methods significantly reduce estimation errors compared to the original approach. Furthermore, the effectiveness and advantages of the proposed methods are validated through both numerical studies and real data analysis.  
	
\noindent \textbf{Keywords:} Statistical simulation; sampling importance resampling; antithetic sampling; Latin hypercube sampling; variance reduction.
\end{abstract}

\section{Introduction}
\label{sec:intro}
The Monte Carlo (MC) method is a widespread simulation technique to validate the efficiency and effect of the statistical theory and method \citep{ross2022}. The basic idea of the MC method is to simulate random sampling using computer software and then analyze statistical problems by collecting sample values generated through simulation. 
It is well-known that the estimators obtained using the MC method often have a large variance. As a result, a series of variance reduction techniques for the MC method have been proposed, such as antithetic variable method, control variable method, Latin hypercube sampling method, importance sampling method, and so on \citep{givens2013,robert2005}. Among them, importance sampling has gained much popularity \citep{tokdar2010}. The basic idea behind importance sampling is to introduce a proposal distribution and sample from it, thereby giving greater ``importance" to the parts of the original distribution that have a significant contribution. Furthermore, some modified methods have also been proposed, such as adaptive importance sampling, sequential importance sampling, and sampling importance resampling.

In Bayesian statistics, one often needs to sample from a target distribution whose regularization constant is unknown or is too complex to compute. The sampling importance resampling (SIR) method proposed by \cite{rubin1987} is one of the efficient methods to deal with  this sampling problem. The process of this method is to generate random samples from the introduced proposal distribution to construct a sample pool, and then randomly resample from the sample pool with a certain importance weight, hence the final obtained samples are  asymptotically i.i.d from the target distribution. In this method, the proposal distribution is usually given and easy to implement. The importance weight measures the difference between the  proposal distribution and the target distribution, which can make samples from the target distribution appeal with higher frequencies in the resampling process.  

In order to improve the performance and applicability of the SIR method, some modified methods have been proposed. For example, \cite{givens1996} proposed the adaptive SIR method in which the sampling process is repeated many cycles, and for each cycle, the proposal distribution is updated through the kernel density estimator. \cite{skare2003} proposed a simple adjustment of the sample probabilities for the SIR algorithm which results in a faster convergence.  \cite{perez2005} proposed two quasi-Monte Carlo versions of the SIR method, that is, the MC points (simple random samples) involved in the SIR method are replaced with the quasi-Monte Carlo points (such as low discrepancy sequence and number-theoretic net \citep{fang1994}), which can lead to  more representative target distribution samples than SIR method. And the convergence of the quasi-random SIR method has been proved by \cite{vandewoestyne2010}. Recently, \cite{ning2020} embedded the randomized quasi-Monte Carlo technique into the SIR method, which introduces randomness into the quasi-random SIR method. Following the idea of SIR method, \cite{yi2023} proposed the global likelihood sampler that is independent of the proposal distributions. In this work, in order to reduce variance and to improve the sampling effect of the SIR method, two variance reduction techniques are embedded into the resampling process of SIR method. The modified sampling methods can get more accurate estimators in statistical inference problems compared with the original  method. 

The rest of this paper is organized as follows. In Section \ref{sec:sir}, the original SIR method is introduced briefly. Then the sampling process and properties of the proposed methods are demonstrated in Section \ref{sec:new}. The numerical simulations and a real data case are used to validate the effect of the proposed methods in Section \ref{sec:num} and Section \ref{sec:data}. Section \ref{sec:conc} concludes this paper.  

\section{Sampling importance resampling}
\label{sec:sir}
Suppose that the density function of a random variable $\bm X$ is $f(\bm x)$, then the expectation of the random function $h(\bm X)$ can be denoted as $\mu:=E[h(\bm X)]=\int h(\bm x)f(\bm x)d \bm x $. If $\bm x_1, \bm x_2, \cdots, \bm x_n$ are i.i.d. samples drawn from $f(\bm x)$, according to the law of large number, the expectation can be approximated by the sample mean, i.e., 
\begin{equation}\label{eq:mc}
	\hat{\mu}_{MC}=\frac{1}{n}\sum_{i=1}^nh(\bm x_i)\rightarrow \int h(\bm x)f(\bm x)d \bm x=\mu, \quad n\rightarrow\infty. 
\end{equation}
Thus $\hat{\mu}_{MC}$ is usually deemed as the MC estimator of $\mu$ when $\bm x_i, i=1,\cdots, n$ are MC samples (i.e., simple random samples).  

In the process of MC estimator, a crucial issue is to generate samples $\bm x_1, \bm x_2, \cdots, \bm x_n$ from the density function $f(\bm x)$. This can be achieved using methods such as the inverse-transformation method. This sampling method usually requires the target distribution to have an explicit density function. However, in statistics, particularly in Bayesian inference, the target density function is often expressed as $f(\bm x)=c\cdot p(\bm x)$ where $c$ is an unknown constant or with an extremely complex expression, in such cases, directly sampling from the target distribution can be prohibited. 

Thus the sampling importance resampling (SIR) method is one of the efficient methods that can deal with the above issue. The basic process of this method can be divided into the following three steps: 
\begin{itemize}
	\item \textbf{Step 1 (Sampling)}: Generating $N$ random samples $\bm x_1, \bm x_2, \cdots, \bm x_N$ from the proposal distribution $g(\bm x)$;
	\item \textbf{Step 2 (Weighting)}: Calculating the importance weight: 
	$$\omega_i=\frac{p(\bm x_i)/g(\bm x_i)}{\sum_{j=1}^Np(\bm x_j)/g(\bm x_j)}, i=1, \cdots, N;
	$$
	\item \textbf{Step 3 (Resampling)}: Sampling $n$ samples from the sample pool $\{\bm x_i, i=1, \cdots, N\}$ with the corresponding weight $\omega_i, i=1,\cdots, N$, denoting as $\{\bm x_1^*, \cdots, \bm x_n^*\}$.
\end{itemize}
According to \cite{givens2013}, the samples $\{\bm x_1^*, \cdots, \bm x_n^*\}$ generating through the SIR method asymptotically converge to target distribution $f(\bm x)$ when $N\rightarrow\infty$ and $n$ is fixed.  Hence, once the samples $\{\bm x_1^*, \cdots, \bm x_n^*\}$  are obtained, the MC integration in \eqref{eq:mc} can be calculated through the SIR samples instead of the MC samples, and the corresponding estimator of integral $\mu$  can be denoted as $\hat{\mu}_{SIR}$.

 
 
\section{Two modified methods}
\label{sec:new}
In nature, step 3 of the SIR method illustrated in Section \ref{sec:sir} is to generate $n$ random samples from the following multinomial distribution $M(N; \omega_1,\omega_2,\cdots,\omega_N)$:
\begin{equation}\label{eq:mutinomial}
	\left(\begin{matrix}
	\bm x_1 & \bm x_2 & \cdots & \bm x_N \\
	\omega_1 & \omega_2 & \cdots & \omega_N
\end{matrix} \right).
\end{equation}
For this discrete multinomial distribution, simple random sampling method can be implemented to generate the samples via the inverse-transformation operator. That is, firstly generating MC samples from the uniform distribution, then the samples from this multinomial distribution can be found via transforming the MC samples with the inverse cumulative distribution function. However, the simple random sampling above will result in a large variance \citep{ross2022,robert2005, fang1994}. 
  
\subsection{SIR with antithetic samples}
In this work, we first propose a modified SIR method to reduce the variance with the help of the antithetic variable method. The antithetic variable method is one of the widespread variance reduction techniques \citep{robert2005}. The basic idea of this method is to generate another estimator $\hat{\mu}_2$ that is identically distributional but negatively correlational with the given MC estimator $\hat{\mu}_1$ of the integral $\mu$. Then by averaging these two estimators, the antithetic estimator is obtained. That is, the key issue of the antithetic variable method is to generate antithetic samples corresponding to the MC samples to estimate the integral $\mu$. 

Through embedding antithetic sampling into the process of the SIR method, a new sampling method, named as ``antithetic sampling importance resampling" method (denoted as ``Anti-SIR" for simplicity), is proposed. The first two steps of the Anti-SIR method are the same as the SIR method. Step 3 of the new method is to generate the antithetic samples from the sample pool $\{\bm x_i, i=1, \cdots, N\}$. Specifically, 
\begin{itemize}
	\item \textbf{Step 3' (Antithetic resampling)}: Sampling $n$ antithetic samples $\{ \bm z_1, \bm z_2, \cdots, \bm z_n \}$ from the sample pool $\{\bm x_i, i=1, \cdots, N\}$ by using the antithetic variable method. 
\end{itemize}
When the resampling size $n$ is even, $\bm z_i$ and $\bm z_{n/2+i}$ are the corresponding antithetic samples for $i=1, \cdots, n/2$; When $n$ is odd, one can first sample $n-1$ antithetic samples and then randomly sample the last one.  

\subsubsection{The sampling details}
In order to demonstrate the process of Anti-SIR method in detail, the size of antithetic samples is assumed to be $2n$ in the following analysis. The key issue of Anti-SIR method is to generate samples $\{\bm z_1, \cdots, \bm z_n, \widetilde{\bm z}_1, \cdots, \widetilde{\bm z}_n\}$ from the sample pool $\{\bm x_i, i=1, \cdots, N \}$, where $\widetilde{\bm z}_i$ is the antithetic sample correspond to $\bm z_i$ for $i=1,\cdots, n$. Thus, firstly $n$ random numbers $u_1, \cdots, u_n$ following the uniform distribution U$[0,1]$ are generated. Through the inverse-transformation method, we can get 
$$
\bm z_i=\begin{cases}
	\bm x_1, & \text{if } u_i\leq \omega_1, \\
	\bm x_2, & \text{if } \omega_1 < u_i\leq \omega_1+\omega_2,\\
	\cdots \\
	\bm x_N, & \text{if } \sum_{k=1}^{N-1} \omega_k < u_i\leq 1,
\end{cases}
\quad i=1,\cdots,n.
$$
Then let $\widetilde{u}_i=1-u_i, i=1,\cdots, n$, the antithetic samples corresponding to these $n$ samples $\bm z_1, \cdots, \bm z_n$ can be generated as follows: 
$$
\widetilde{\bm z}_i=\begin{cases}
	\bm x_1, & \text{if } \widetilde u_i\leq \omega_1, \\
	\bm x_2, & \text{if } \omega_1 <\widetilde u_i\leq \omega_1+\omega_2,\\
	\cdots \\
	\bm x_N, & \text{if } \sum_{k=1}^{N-1}\omega_k < \widetilde u_i\leq 1,
\end{cases}
\quad i=1,\cdots,n.
$$
Following the framework of SIR method, the obtained samples $\{\bm z_1, \cdots, \bm z_n, \widetilde{\bm z}_1, \cdots, \widetilde{\bm z}_n\}$ via Anti-SIR method are distributed from the target distribution $f(\bm x)$ asymptotically. 

\subsubsection{Theoretical properties}
In this subsection, the statistical properties of $2n$ samples $\{\bm z_1, \cdots, \bm z_n, \widetilde{\bm z}_1, \cdots, \widetilde{\bm z}_n\}$ generating by Anti-SIR method are discussed. 



\begin{Thm}\label{theorem:antisir}
When using SIR and Anti-SIR methods to estimate the integral $\mu=\int h(\bm x)f(\bm x)d \bm x$ respectively, suppose the sample size to be $2n$, the estimator corresponding to SIR method can be written as $\hat{\mu}_{SIR}=\frac{1}{2n}\sum_{i=1}^{2n}h(\bm x_i^*)$ and the estimator corresponding to Anti-SIR method can be written as $\hat{\mu}_{ASIR}=\frac{1}{2n}\sum_{i=1}^n[h(\bm z_i)+h(\tilde{\bm z}_i)]$, we have
\begin{itemize}
		\item[(\romannumeral1)] $E(\hat{\mu}_{ASIR})=E(\hat{\mu}_{SIR})$;
		\item[(\romannumeral2)] $Var(\hat{\mu}_{ASIR})\leq Var(\hat{\mu}_{SIR})$.
\end{itemize}
\end{Thm}

Theorem \ref{theorem:antisir} shows that Anti-SIR method has the same bias as SIR method and both of them are unbiased estimators of integral $\mu$. In addition, by embedding the antithetic sampling method into the original SIR method, the variance of the new estimator $\hat{\mu}_{ASIR}$ can be reduced. 

\subsection{Latin hypercube sampling (LHS)-based SIR}
It is well-known that Latin hypercube sampling (LHS) is another popular variance reduction technique proposed by \cite{mckay1979}, which is widely used in computer experiments \citep{santner2018} and MC simulations \citep{ross2022}. The basic idea of LHS method is to divide the domain along with each dimension into sub-intervals with equal marginal probability, and sample once from each stratum. In this way, the integration estimator via LHS can possess lower variance than that via simple random sampling. 

Hence, in this subsection, another variance-reduced SIR method based on Latin hypercube sampling is proposed, denoted as ``LHS-SIR" for convenience. Similar to Anti-SIR method, the only difference between this modified method with the original SIR method is the resampling operation. Specifically,        
\begin{itemize}
	\item \textbf{Step 3'' (Latin hypercube resampling)}: Generating $n$ Latin hypercube samples $\{\bm{y}_1, \bm{y}_2, \cdots, \bm{y}_n\}$ from the sample pool $\{\bm x_i, i=1,\cdots, N\}$ by using the LHS method. 
\end{itemize}
Through this resampling process, it can be observed that the variance for the integral estimator via the original SIR method can be reduced. 

\subsubsection{The sampling details}
The pivotal element in the proposed LHS-SIR method is to generate samples from the multinomial distribution \eqref{eq:mutinomial} via the Latin hypercube sampling technique. The detailed procedures are demonstrated as following three steps: 
\begin{itemize}
	\item[1.] Dividing the interval $[0, 1]$ into $n$ strata with equidistance, i.e., 
	$$\left[0,\frac{1}{n}\right], \left[\frac{1}{n}, \frac{2}{n}\right], \cdots, \left[\frac{n-1}{n}, 1\right];$$
	\item[2.] Generating a random number $v_i$ in each sub-interval, i.e., 
	$$v_i\sim U\left[\frac{i-1}{n}, \frac{i}{n}\right], i=1,\cdots,n;$$
	\item[3.] Finding $\bm y_i$ corresponding to each $v_i$ via the inverse-transformation method, i.e.,
$$
\bm y_i=\begin{cases}
	\bm x_1, & \text{if } v_i\leq \omega_1, \\
	\bm x_2, & \text{if } \omega_1 < v_i\leq \omega_1+\omega_2,\\
	\cdots \\
	\bm x_N, & \text{if } \sum_{k=1}^{N-1}\omega_k < v_i\leq 1,
\end{cases}
\quad i=1,\cdots,n.
$$ 
\end{itemize}
Hence based on the framework of SIR method, the obtained samples $\{\bm y_i, i=1,\cdots,n\}$ via LHS-SIR method asymptotically converge to the target distribution $f(\bm x)$. 

\subsubsection{Theoretical properties}
In this subsection, the statistical properties of $n$ samples $\{\bm y_1, \bm y_2, \cdots, \bm y_n\}$ generated by LHS-SIR method are discussed. Similar to Anti-SIR method, LHS-SIR method also reduces the variance of the estimation by generating the samples with a negative correlation. Then, we have the following result.



\begin{Thm}\label{theorem:lhssir}
When using SIR and LHS-SIR methods to estimate the integral $\mu=\int h(\bm x)f(\bm x)d \bm x$ respectively, suppose the sample size to be $n$ here, the estimator corresponding to SIR method can be written as $\hat{\mu}_{SIR}=\frac{1}{n}\sum_{i=1}^{n}h(\bm x_i^*)$ and the estimator corresponding to LHS-SIR method can be written as $\hat{\mu}_{LSIR}=\frac{1}{n}\sum_{i=1}^nh(\bm y_i)$, we have
\begin{itemize}
		\item[(\romannumeral1)] $E(\hat{\mu}_{LSIR})=E(\hat{\mu}_{SIR})=\mu$;
		\item[(\romannumeral2)] $Var(\hat{\mu}_{LSIR})\leq Var(\hat{\mu}_{SIR})$.
\end{itemize}
\end{Thm}

The results of Theorem \ref{theorem:lhssir} are similar to that of Theorem \ref{theorem:antisir}. By embedding the Latin hypercube resampling method into the original SIR method, the variance of the estimation can be reduced. In the next sections, we will evaluate the performances of these methods by extensive numerical studies.

\section{Numerical studies}
\label{sec:num}
In this section, some numerical simulations with different dimensions are conducted to validate the performance of the modified SIR method. 

Suppose $\bm{X}=(x_1,x_2,\cdots,x_d)^\top$ is the $d$-dimensional random variable and the density function is $f(\bm{x}),\bm{x}\in\mathcal{X}$, where $\mathcal{X}$ is the corresponding sample space. Here, we want to estimate the expectation of random variable $\bm X$:
\begin{equation}\label{eq:expect}
	\mathrm{E}(\bm{X})=\int_{\mathcal{X}}\bm{x}f(\bm{x})\mathrm{d}\bm{x},
\end{equation}
by the corresponding samples mean with $n$ samples from target distribution $f(\bm x)$ generated through Anti-SIR and SIR methods. 
In order to alleviate the randomness in the sampling process, the operation is replicated $K$ times and the obtained  samples each time are denoted as $\bm{x}_{i}^k,i=1,2,\cdots,n,k=1,2,\cdots,K$. Then $\mathrm E(\bm{X})$ can be estimated by 
$$\hat{\mathrm{E}}_k(\bm{X})=\dfrac{1}{n}\sum_{i=1}^{n}\bm{x}_i^k, k=1,2,\cdots,K,
$$ 
and the mean square error (MSE) of these estimations is given by
\begin{equation}\label{eq:mse}
	\text{MSE}=\dfrac{1}{K}\sum_{k=1}^K\left[\hat{\mathrm{E}}_k(\bm{X})-\overline{\hat{\mathrm{E}}_k(\bm{X})}\right]^2,
\end{equation}
where $\overline{\hat{\mathrm{E}}_k(\bm{X})}=\dfrac{1}{K}\sum_{k=1}^K\hat{\mathrm{E}}_k(\bm{X})$. If the true value of expectation \eqref{eq:expect} is known in advance, it could be used to replace the term $\overline{\hat{\mathrm{E}}_k(\bm{X})}$ in \eqref{eq:mse}. The smaller the MSE value, the more accurate the estimation, indicating the better the performance of the sampling method.

\textbf{Example 1}: Univariate distributions. 

In this example, several cases in dimension 1 are considered. The target distributions in this example are Beta distribution, normal distribution, $t$-distribution, and $F$-distribution; and the proposal distributions are uniform distribution, logistic distribution, Cauchy distribution, and inverse-Gaussian distribution. The sample size chosen for all the cases is $N=20000$ and the resampling size is $n=1000$. All the simulations are repeated $K=1000$ times. The results obtained by SIR, Anti-SIR, and LHS-SIR methods are listed in \autoref{tab:mean}. It can be shown clearly that both the proposed Anti-SIR and LHS-SIR methods perform better than the original SIR method in all cases. These two modified methods can reduce the mean squared errors for the SIR method with the help of antithetic resampling and Latin hypercube resampling. However, in this example, neither of these two modified methods is consistently better than the other. 

\begin{table}[h]
\centering
\caption{MSEs in the estimation of the mean}
\label{tab:mean}
	\begin{tabular}{ccccc}
	\toprule
	$f(x)$ & $g(x)$ & SIR & Anti-SIR & LHS-SIR \\ \midrule
	Beta(2, 3) & U(0, 1) & $4.190\times 10^{-5}$ & $4.007\times 10^{-5}$ & $3.990\times 10^{-5}$ \\
	Beta(0.9, 0.9) & U(0, 1) & $1.037\times 10^{-4}$ & $9.613\times 10^{-5}$ & $9.023\times 10^{-5}$ \\
	N(0, 1) & Log(0, 1) & $1.144\times 10^{-3}$ & $ 1.070\times 10^{-3}$ & $1.053\times 10^{-3}$  \\
	N(0, 1) & C(0, 1) & $1.136\times 10^{-3}$ & $9.898\times 10^{-4}$ & $1.069\times 10^{-3}$\\
	$t$(2, 0, 1) & C(0, 1) & $2.377\times 10^{-2}$  & $1.478\times 10^{-2}$ & $1.545\times 10^{-2}$ \\
	$F$(10, 6) & IG(1, 1) & $3.512\times 10^{-3}$ & $2.963\times 10^{-3}$ & $3.091\times 10^{-3}$ \\ \bottomrule
	\end{tabular}
\end{table}

\textbf{Example 2}: Multivariate distribution. 

In this example, the target distribution is the Kotz-type distribution which is a generalization of the multivariate normal distribution \citep{kotz1975,fang1990}. Suppose the $d$ dimensional random variable $\bm X$ is following the Kotz-type distribution with the density function: 
$$f(\bm{x})=C_p|\bm\Sigma|^{-1/2}\left[(\bm{x}-\bm\beta)^\top\bm\Sigma^{-1}(\bm{x}-\bm \beta)\right]^{M-1}\exp\left\{-r\left[(\bm{x}-\bm \beta)^\top\bm\Sigma^{-1}(\bm{x}-\bm\beta)\right]^s\right\},
$$
where $r>0,s>0,2M+d>2$, $\bm\Sigma$ is the $d\times d$ positive-definite dispersion matrix, $\bm\beta=(\beta_1, \cdots, \beta_d)^\top$ is the mean vector, and $C_p$ is the normalizing constant. Here, we consider a 4-dimensional Kotz-type distribution with the parameters: $r=0.5,s=2,M=3,\bm \beta =(0,0,0,0)^\top$, and 
$$\Sigma=\begin{pmatrix}5.3 & 0.0 & 0.0 & -0.2\\ 0.0 & 4.0 & -0.4 & 0.3\\ 0.0 & -0.4 & 6.8 & 0.0\\ -0.2 & 0.3 & 0.0 & 9.0
\end{pmatrix}.
$$ 

In order to generate samples from the Kotz-type distribution through the SIR and Anti-SIR methods, the multivariate normal distribution with mean vector $\bm \beta_g=\bm \beta $ and covariance matrix $\bm \Sigma_g=\bm \Sigma$
is taken as the proposal distribution. We set $N=2000,n=400,K=1000$. The samples drawn by the SIR and Anti-SIR methods are used to estimate the expectation of the Kotz-type distribution. \autoref{tab:kotz}  lists the MSEs of the estimated components of the mean and the overall MSEs (OMSE). It can be shown that for the MSEs of each component and the overall MSE, both Anti-SIR and LHS-SIR methods are superior to SIR method. Hence, the proposed methods in our work are effective sampling methods in this example. In this case, LHS-SIR method performs a little better than Anti-SIR method. 

\begin{table}[h]
\centering
\caption{MSEs and OMSE in the estimation of the mean vector}
	\label{tab:kotz}
	\begin{tabular}{cccccc}
	\toprule
	Methods & MSE ($\beta_1$) & MSE ($\beta_2$)  & MSE ($\beta_3$) & MSE ($\beta_4$) & OMSE  \\ \midrule
	SIR & 0.008897 & 0.006810 & 0.01227 & 0.01561 & 0.04359 \\
	Anti-SIR & 0.008950 & 0.006707 & 0.01170 & 0.01526 & 0.04262 \\
	LHS-SIR & 0.006330 & 0.004795 &  0.009020 & 0.01149 & 0.03163 \\
	 \bottomrule
	\end{tabular}
\end{table}

\section{Real data analysis}
\label{sec:data}
\begin{figure}[h]
    \centering
	\includegraphics[height=8cm, width=10cm]{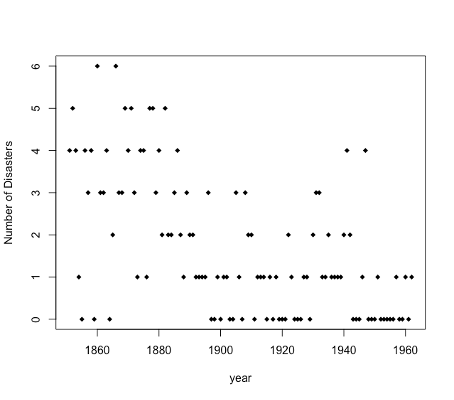}
	\caption{The number of coal-mining disasters per year between 1851 and 1962}
	\label{fig:mining}
\end{figure}

In this section, a real data analysis case is used to illustrate the performance in Bayesian inference for our proposed method. Here, the data on the number of coal-mining disasters per year between 1851 and 1962 is plotted in \autoref{fig:mining}. It can be shown that the rate of accidents per year appears to decrease around 1900, hence a change-point model is constructed to analyze these data \citep{carlin1992}. 

Let $X_j$ be the number of coal-mining disasters in year $j$, where we denote $j=1$ in the year 1851 and index each year thereafter, i.e., $j=112$ in the year 1962. Hence suppose the change-point occurs after the $\theta$year in the series, $\theta\in\{1,\cdots,111\}$,  then the change-point model can be expressed as 
$$ X_1,\cdots,X_{\theta}\sim\mathrm{Poisson}(\lambda_1),
$$
and 
$$X_{\theta+1},\cdots,X_{112}\sim\mathrm{Poisson}(\lambda_2), 
$$
where $\theta, \lambda_1$ and $\lambda_2$ are the unknown parameters and the estimator of $\theta$ is our primary interest.  

In order to estimate the parameters $\Theta=\{\theta, \lambda_1, \lambda_2\}$ for this model, suppose the prior distribution for $\Theta$ is $\pi(\Theta)$, and the likelihood function is: 
\begin{equation}\label{eq:like}
	 \ell(\Theta|\bm{x})=\prod_{i=1}^{\theta}\dfrac{\lambda_1^{x_i}}{x_i!}\mathrm{e}^{-\lambda_1}\prod_{j=\theta+1}^{112}\dfrac{\lambda_2^{x_j}}{x_j!}\mathrm{e}^{-\lambda_2}.
\end{equation}
Then with prior and the likelihood \eqref{eq:like}, the posterior for $\Theta$ can be written as $f(\Theta| \bm x)\propto \ell(\Theta|\bm{x})\pi(\Theta)$. In this work, the SIR, Anti-SIR, and LHS-SIR methods are used to draw samples from this posterior distribution $f(\Theta| \bm x)$. That is, the posterior is the target density. Let the prior distribution $\pi(\Theta)$ sever as the proposal distribution, i.e., sampling $\Theta_1, \cdots, \Theta_N$ i.i.d. from $\pi(\Theta)$, then the $i$th unstandardized weight equals $ \ell(\Theta_i|\bm{x})$, $i=1,\cdots, N$. Then the i.i.d resampling, antithetic resampling, and Latin hypercube resampling samples from $\Theta_1, \cdots, \Theta_N$ can be employed to estimate the unknown parameters respectively.  

Here, we set two different priors for a Bayesian analysis of this model: 
\begin{itemize}
	\item Case 1: $\theta$ is from the discrete uniform distribution on $\{1, 2, \cdots,111\}$, $\lambda_i|a_i\sim\mathrm{Gamma}(3, a_i)$ and $a_i\sim\mathrm{Gamma}(10, 10)$ for $i=1,2$;
	\item Case 2: $\theta$ is from the discrete uniform distribution on $\{1, 2, \cdots,111\}$, $\lambda_2=\alpha\lambda_1$, $\lambda_1|a\sim\mathrm{Gamma}(3, a)$, $a\sim\mathrm{Gamma}(10, 10)$  and $\log\alpha\sim \mathrm{U}(\log 1/8, \log 2) $.
\end{itemize}   
where ``Gamma" represents the Gamma distribution and ``U" is the uniform distribution. The sample size chosen for these two cases is $N=5000$ and the resample size is $n=2000$. Repeating the sampling processes $K=1000$ times, the results about the estimated parameters are listed in \autoref{tab:coal}. It can be shown that in both cases, the estimated parameters through SIR, Anti-SIR, and LHS-SIR methods are consistent on average. However, the standard deviations obtained by Anti-SIR and LHS-SIR methods are a little less than those corresponding to SIR method. That is, these two proposed methods scan provide more robust results than that of SIR method in statistical inference. In addition, LHS-SIR method produces better performance than Anti-SIR method except for the estimation of parameter $\lambda_1$ in case 1.    

\begin{table}[h]
\centering
\begin{threeparttable}
	\caption{The estimated parameters of the change-point model}
	\label{tab:coal}
	\begin{tabular}{ccccc}
	\toprule
	 \multirow{2}*{Cases} &  \multirow{2}*{Methods} & \multicolumn{3}{c}{Parameters}\\ \cline{3-5}
	 & & $\theta$ & $\lambda_1$ & $\lambda_2$ \\ \midrule
	 \multirow{2}*{1} & SIR & 39.81 (0.8862) & 3.122 (0.1103) & 0.9569 (0.04513) \\ 
	 & Anti-SIR & 39.81 (0.8831) & 3.121 (0.1101) & 0.9569 (0.04503) \\
	 & LHS-SIR & 39.81 (0.8828) & 3.122 (0.1103) & 0.9569 ( 0.04494) \\
	  \midrule
	     \multirow{2}*{2} & SIR & 39.97 (0.5880) & 3.116 (0.07230) & 0.9243 (0.02926) \\ 
	 & Anti-SIR & 39.97 (0.5876) & 3.116 (0.07198) & 0.9241 (0.02915) \\ 
	 & LHS-SIR & 39.98 (0.5850) & 3.116 (0.07173) & 0.9242 ( 0.02904) \\
   \bottomrule
	\end{tabular}
	\begin{tablenotes}[para]
		$^*$ The results are listed as the form  ``mean (standard deviations)".
	\end{tablenotes}
\end{threeparttable}
\end{table}

\section{Conclusion} 
\label{sec:conc}
In this paper, two modified SIR methods, namely Anti-SIR and LHS-SIR methods, are proposed based on the variance reduction techniques. The core step is to replace the simple random resampling from the sample pool $\{\bm x_1, \cdots, \bm x_N\}$ with the antithetic resampling and Latin hypercube resampling respectively. The variance reduction effects for these two proposed methods are demonstrated in both theoretical and numerical aspects. It can be observed that the new proposed methods can produce lower estimation errors than the original SIR method. In addition, from numerical results, LHS-SIR method performs a little better than Anti-SIR method, especially in the high-dimensional case.

\section*{Acknowledgments}
This work was supported by the National Natural Science Foundation of China [grant numbers: 12371261 and 12171187] and the Humanities and Social Science Foundation of Ministry of Education of China [grant number: 23YJC910004].

\section*{Declarations}
The authors have no relevant financial or non-financial interests to disclose.

\bibliographystyle{agu04}
\bibliography{ref}

\appendix
\section{The proof of the theorems} 
\subsection{The proof of Theorem 1}
In order to prove the theorem 1, a significant lemma is introduced.

\begin{Lem}[Corollary in \cite{ross2022}]\label{lemma:ross}
	If $l(x_1, \cdots, x_m)$ is a monotone function of each of its arguments, then, for a set $U_1, \cdots, U_m$ of independent random numbers drawn from the uniform distribution,
	\[
	Cov(l(U_1, \cdots, U_m), l(1-U_1, \cdots, 1-U_m)) \leq 0.
	\]
\end{Lem}
This lemma indicates that the samples obtained by the antithetic sampling method are negatively correlated. This property plays a key role in reducing the variance of samples by the Anti-SIR method. Formally, Theorem 1 is proved as follows.

\textit{Proof:}
	(\romannumeral1)  For SIR method, according to page 183 of \cite{givens2013},  
	$$ E(\hat{\mu}_{SIR})=E\left\{\frac{1}{2n}\sum_{i=1}^{2n}h(\bm x_i^*) \right\}=E(h(\bm x_1^*)).
	$$
	For Anti-SIR method, antithetic samples $\{\bm z_1, \cdots, \bm z_n, \tilde{\bm z}_1, \cdots, \tilde{\bm z}_n\}$, $\bm z_i$ and $\tilde{\bm z_i}$ are correlated, but pairs $(\bm z_i, \tilde{\bm z_i})$, $i=1,\cdots,n,$ are independent, hence
	$$ E(\hat{\mu}_{ASIR})=E\left\{\frac{1}{2n}\sum_{i=1}^n[h(\bm z_i)+h(\bm  \tilde{z}_i)] \right\}=\frac{1}{2}E[h(\bm z_1)+h(\tilde{\bm z}_1)]=E(h(\bm z_1))=E(\hat{\mu}_{SIR}).
	$$
	
	(\romannumeral2) From (\romannumeral1), we can have 
	 $$ Var(\hat{\mu}_{SIR})=Var\left\{\frac{1}{2n}\sum_{i=1}^{2n}h(\bm x_i^*) \right\}=\frac{1}{2n}Var(h(\bm x_1^*)).
	$$
	Then,  
	\begin{equation*}
			\begin{split}
			Var(\hat{\mu}_{ASIR})&=Var\left\{\frac{1}{2n}\sum_{i=1}^n[h(\bm z_i)+h(\bm  \tilde{z}_i)] \right\}=\frac{1}{4n}Var[h(\bm z_1)+h(\bm  \tilde{z}_1)]\\
			&=\frac{1}{4n}[Var(h(\bm z_1))+Var(h(\bm  \tilde{z}_1))+2Cov(h(\bm z_1), h(\bm  \tilde{z}_1))]\\
			&=\frac{1}{2n}Var(h(\bm z_1)+\frac{1}{2n}Cov(h(\bm z_1), h(\bm  \tilde{z}_1))\\
			&=Var(\hat{\mu}_{SIR})+\frac{1}{2n}Cov(h(\bm z_1), h(\bm  \tilde{z}_1).
	\end{split}
	\end{equation*}
According to Lemma \ref{lemma:ross}, because the inverse cumulative distribution function of multinomial distribution is monotone, $Cov(h(\bm z_1), h(\bm  \tilde{z}_1))\leq0$. Hence, $Var(\hat{\mu}_{ASIR})\leq Var(\hat{\mu}_{SIR})$. The proof is finished. 

\subsection{The proof of Theorem 2}
\textit{Proof:}
	(\romannumeral1) Because LH samples $\{\bm y_i, i=1,\cdots,2\}$ are negatively correlated each other, we have 
	$$ E(\hat{\mu}_{LSIR})=E\left(\frac{1}{n}\sum_{i=1}^nh(\bm y_i)\right)=E(h(\bm y_1))=E(\hat{\mu}_{SIR}).
	$$
	
	(\romannumeral2) From (\romannumeral1), we can get that 
	\begin{equation*}
		\begin{split}
			Var(\hat{\mu}_{LSIR})&=Var\left(\frac{1}{n}\sum_{i=1}^nh(\bm y_i)\right)\\
			&= \frac{1}{n^2}\left[\sum_i Var(h(\bm y_i))+2\sum_{i,j}Cov(h(\bm y_i), h(\bm y_j))\right] \\
			&=\frac{1}{n}Var(h(\bm y_1))+\frac{2}{n^2}\sum_{i,j}Cov(h(\bm y_i), h(\bm y_j))\\
			&=Var(\hat{\mu}_{SIR})+\frac{2}{n^2}\sum_{i,j}Cov(h(\bm y_i), h(\bm y_j)).
		\end{split} 
	\end{equation*}
	According to the results in \cite{mckay1979}, we have $Cov(h(\bm y_i), h(\bm y_j))\leq0$ for any $i, j\in \{1,\cdots,n\}$. Hence, we can obtain $Var(\hat{\mu}_{LSIR}) \leq Var(\hat{\mu}_{SIR})$. The proof is finished.
\end{document}